\documentstyle[12pt]{article}
\include{epsf}
\def\lo{\langle 0 |}
\def\ro{ | 0 \rangle }

\def\gmf{\gamma _{5}}

\def\la{\langle }
\def\ra{ \rangle }

\newcommand{\beq}{\begin{equation}}
\newcommand{\eeq}{\end{equation}}
\newcommand{\bea}{\begin{eqnarray}}
\newcommand{\eea}{\end{eqnarray}}

\begin{document}
                                        \begin{titlepage}
\begin{flushright}
hep-ph/9803301
\end{flushright}
\vskip1.8cm
\begin{center}
{\LARGE
 Anomalous Effective Lagrangian  \\       
\vskip0.7cm
and $ \theta $ Dependence in QCD at Finite $ N_c $    \\
%\vskip1.0cm
 %and 4D Gluodynamics    
}         
\vskip1.5cm
 {\Large Igor~Halperin} 
and 
{\Large Ariel~Zhitnitsky}
\vskip0.5cm
        Physics and Astronomy Department \\
        University of British Columbia \\
 6224 Agricultural Road, Vancouver, BC V6T 1Z1, Canada \\ 
        {\small e-mail: 
higor@physics.ubc.ca \\
arz@physics.ubc.ca }
\vskip1.0cm
{\Large Abstract:\\}
\end{center}
\parbox[t]{\textwidth}{
We generalize the large $N_c $ Di Vecchia-Veneziano-Witten (VVW)
effective chiral Lagrangian to the case of finite $ N_c $
by constructing the anomalous effective Lagrangian for QCD. The latter
is similar to its SUSY counterpart, and has a holomorphic structure.
The VVW construction 
is then recovered, along with $ 1/N_c $ corrections, after integrating 
out the heavy ``glueball" fields. A new mass formula for $ \eta' $ meson
in terms of QCD condensates is obtained. The picture of 
$ \theta $ dependence in QCD for finite $ N_c $ is more complicated
than the one predicted by the large $ N_c $ approach.   
}

\vspace{1.0cm}

                                                \end{titlepage}

{\bf 1.} There exist two different definitions of an 
effective Lagrangian 
in quantum field theory. One of them is Wilsonian effective 
Lagrangian describing the low energy dynamics of the lightest particles
in the theory. In QCD, this is implemented by effective chiral
Lagrangians (ECL's) for the pseudoscalar mesons, which are 
essentially fixed by the global non-anomalous $ SU(N_f) \times 
SU(N_f) $ and (for large $ N_c $) anomalous $U(1) $ chiral symmetries.
Another type of the effective Lagrangian (action) is defined as  
the Legendre transform of the generating functional
for connected Green functions. This 
object is relevant for addressing the vacuum properties of the 
theory in terms of vacuum expectation values (VEV's) of composite
operators, as they should minimize the effective action.
Such an approach is suitable for the study of the dependence 
of the QCD vacuum on external parameters, such as the light quark masses 
or the vacuum angle $ \theta $. The lowest dimensional
condensates $ \la \bar{\Psi} \Psi \ra , \la G^2 \ra,
\la G \tilde{G} \ra $, which are the most essential for the 
QCD vacuum structure,
are related to the anomalously and explicitly broken conformal 
and chiral symmetries of QCD. Thus, one can study the vacuum of QCD
with an effective Lagrangian realizing at the tree level 
anomalous conformal and chiral Ward identities of the theory.

The utility of such an approach to gauge theories has been recognized 
long ago for supersymmetric models, where anomalous effective
Lagrangians have been found for both the pure gauge case \cite{VY}
and super-QCD \cite{TVY}. A non-supersymmetric analog of 
the Veneziano-Yankielowicz (VY) effective potential \cite{VY}
was suggested only recently \cite{1}. The purpose of this letter
is to extend the construction \cite{1} to the case of 
full QCD with $ N_f $ light flavors and $ N_c $ colors,
i.e. to find an effective Lagrangian (more precisely, its potential
part) realizing the anomalous conformal and chiral 
symmetries of QCD.

The interest in such an effective Lagrangian is two-fold.
First, it provides a generalization of the large $N_c $ 
Di Vecchia-Veneziano-Witten (VVW) ECL \cite{Wit2} (see also \cite{ECL})
for the case of arbitrary
$ N_c $ after integrating out the massive ``glueball" fields.
This helps to better understand the origin of the 
$ \eta' $ mass (the famous U(1) problem). In particular, 
it yields a new mass formula for the $ \eta' $ for finite $ N_c $ in 
terms 
of quark and gluon condensates in QCD (see Eq.(\ref{12}) below).
Secondly, such an effective Lagrangian allows one
to address the problem of $ \theta $ dependence in QCD.
In contrast to the approach of Ref.\cite{Wit2} which deals 
from the very beginning with the light chiral degrees of freedom and 
explicitly incorporates the $ U_{A}(1) $ anomaly without restriction
of the topological charge to integer values, in our method both 
$ U_{A}(1) $ anomaly and topological charge quantization are 
included in the effective Lagrangian framework. 
After the ``glueball" fields are integrated out, the topological
charge quantization still shows up in the limit $ V \rightarrow
\infty $ via the presence of certain 
cusps in the effective potential, which 
are not present in the large $ N_c $ ECL of 
Ref.\cite{Wit2}. Analogous ``glued" effective 
potentials containing cusp
singularities arise in supersymmetric $ N = 1 $ theories when 
quantization of the topological charge is imposed \cite{KS,KKS}. 
As will be discussed below, these modifications are not essential
for the local properties of the effective chiral potential
in the vicinity of the global minimum.
 In this case, the results of Ref.\cite{Wit2} are 
reproduced
along with calculable $ 1/N_c $ corrections. On the other 
hand, for large
values of $ \theta $ and /or $ \phi_i $ our results deviate from those 
of \cite{Wit2}.

{\bf 2.} We start with recalling the construction \cite{1} 
of the anomalous effective potential for pure YM theory (gluodynamics).
It is constructed as the Legendre transform of the generating 
functional for zero momentum correlation functions of the 
marginal operators $  G_{\mu \nu} 
\tilde{G}_{\mu \nu} $ and $ G_{\mu \nu} 
G_{\mu \nu} $ which are related
to the gluon condensate due to conformal anomaly\cite{NSVZ,2}. 
The effective potential
is a  function of effective zero momentum fields $ h , \bar{h} $
which describe the VEV's of the composite 
complex fields 
$ H ,\bar{H} $ :
\beq
\label{1}
\int dx \, h = \la \int dx \, H \ra \; \; , \; \; 
\int dx \, \bar{h} =  \la \int dx \, \bar{H} \ra \; , 
\eeq
where 
\beq
\label{2}
H  = \frac{b_{YM} \alpha_s}{16 \pi} \left( -
G^2  + i \, \frac{2}{b_{YM} \xi_{YM}} \ 
G \tilde{G} \right) \; , \; \bar{H}   
= \frac{b_{YM} \alpha_s}{16 \pi} \left( - 
G^2  - i \, \frac{2}{b_{YM} \xi_{YM}} 
G \tilde{G} \right) \; .
\eeq 
Here $ b_{YM} = (11/3) N_c $ is the first coefficient of the 
Gell-Mann - Low $ \beta $-function for YM theory, and $ \xi_{YM} $
is a generally unknown rational number, $ \xi \equiv q/(2p) $
(here $ p $ and $ q $ are relatively prime integers), 
fixing the topological susceptibility in terms of the gluon condensate
\cite{1}. On general grounds, it follows that $ p = O(N_c) \, , \, q = 
O(N_{c}^0 ) $. (For some plausible line of reasoning leading to
the particular value $ \xi =  4/( 3 b_{YM} ) $, see \cite{3}.) 
The advantage of 
using the combinations (\ref{2})
is in the holomorphic structure of zero momentum correlation functions
of $ H, \bar{H} $ fields \cite{1,3,KZ}. 
As a result, the effective potential also has the holomorphic structure.  
The final answer for the 
improved effective potential (IEP) $ W(h, \bar{h}) $ 
(here 'improved' refers
to the necessity of summation over the integers $n, k $ in 
Eq.(\ref{3}),
see below) reads \cite{1}   
\bea
\label{3}
e^{- i V W(h,\bar{h}) } &=& \sum_{n = - \infty}^{
 + \infty} \sum_{k=0}^{q-1} \exp \left\{ - \frac{i V}{4}
\left( h \, Log \, \frac{h}{C_{YM}} + 
\bar{h} \, Log \, \frac{ \bar{h}}{
\bar{C}_{YM}} \right) \right. \nonumber \\ 
&+& \left. i \pi V \left( k + \frac{q}{p} \,  
\frac{ \theta + 2 \pi n}{ 2 \pi} \right) \frac{h - \bar{h}}{
2 i} \right\} \;  ,
\eea
where    
the constants $ C_{YM}, \bar{C}_{YM} $ can be 
taken real and expressed in terms 
of the vacuum energy in YM theory at $ \theta = 0 $, 
$  C_{YM} =  \bar{C}_{YM} = - 2 e E_{v}^{(YM)}(0) 
= - 2 e \la - b_{YM} \alpha_s/(32 \pi) G^2 \ra $, and $ V $ 
is the 4-volume.
The symbol $ Log $ in 
Eq.(\ref{3}) stands for the principal branch of the logarithm.
The effective potential (\ref{3}) produces an infinite series
of anomalous Ward identities (WI's) and has the 
correct $ 2 \pi $ periodicity in $ \theta $. It 
is suitable for a study of the YM vacuum as described above. 

The double sum over the integers 
$ n , k $ in
Eq.(\ref{3}) appears as a resolution of an ambiguity
of the effective potential as defined from the anomalous WI's.
As was discussed in \cite{1}, this ambiguity is due to the fact
that any particular branch of the multi-valued function
$ h \log (h/c)^{p/q} $, corresponding to some fixed 
values of $ n , k $, satisfies the anomalous WI's.
However, without the summation over the integers $ n , k $ 
in Eq.(\ref{3}), the effective potential would be   
multi-valued and unbounded from below. 
An analogous problem arises with the 
original VY effective Lagrangian. It was cured 
by Kovner and Shifman in \cite{KS}
by a similar prescription of summation over all branches of the 
multi-valued VY superpotential. Moreover, the whole 
structure of Eq.(\ref{3}) is rather similar to that of the (amended)
VY effective potential. Namely, it contains
both the ``dynamical" and ``topological" parts (the 
first and the second terms in the exponent, respectively).
The ``dynamical" part of the effective potential (\ref{3}) 
is similar to the VY \cite{VY}
potential $
\sim S \log (S/ \Lambda)^{N_c} $ (here 
$ S $ is an anomaly superfield), 
while the ``topological" part is 
akin to the improvement \cite{KS} of the VY effective potential. 
Similarly to the supersymmetric case, the infinite sum over $ n $ 
reflects
the summation over all integer topological charges in the 
original YM theory. Thus, the topological charge quantization
appears naturally in the effective Lagrangian framework, and 
leads to the single-valuedness and boundness from below of the
IEP (\ref{3}).

{\bf 3.} We now proceed to the generalization of the 
result (\ref{3}) to 
the case of full QCD with $ N_f $ light flavors and $ N_c $ colors.
In the effective Lagrangian approach, the light matter fields are 
described by the unitary matrix $ U_{ij} $ corresponding to the
$ \gmf $ phases of the chiral condensate: $ \la \bar{\Psi}_{L}^{i} 
\Psi_{R}^{j} \ra 
=  - | \la \bar{\Psi}_{L} \Psi_{R} \ra | \, U_{ij} $ with 
\beq
\label{4}
U = \exp \left[ i \sqrt{2} \, \frac{\pi^{a} \lambda^{a} }{f_{\pi}}  + 
i \frac{ 2}{ \sqrt{N_{f}} } \frac{ \eta'}{ f_{\eta'}}  \right] 
\; \; , \; \; 
U U^{+} = 1 \; ,
\eeq
where $ \lambda^a  $ are the Gell-Mann matrices of $ SU(N_f) $, 
$ \pi^a $ 
is the pseudoscalar octet, and 
$ f_{\pi} = 133 \; MeV $. As is well known \cite{Wit2}, the 
effective potential for the $ U $ field (apart from the mass term) is   
uniquely fixed by the chiral anomaly, and amounts to the substitution 
\beq
\label{5} 
\theta \rightarrow \theta - i \, Tr \, \log U \; 
\eeq
in the topological density term in the QCD Lagrangian.
Note that for spatially independent vacuum fields $ U $ Eq.(\ref{5})
results in the shift of $ \theta $ by a constant.
This fact will be used below. Furthermore, in the sense 
of anomalous conformal Ward identities \cite{NSVZ} QCD reduces to 
pure YM theory when the quarks are ``turned off" with the 
simultaneous substitution $ \la G^2 \ra_{QCD} \rightarrow 
\la G^2 \ra_{YM} $ 
% (or $ \Lambda_{QCD} \rightarrow \Lambda_{YM} $)
and $ b \equiv b_{QCD} \rightarrow b_{YM} $.
Analogously,  
an effective Lagrangian for QCD should transform to
that of pure YM theory when the chiral fields $ U $ are ``frozen".    
Its form is thus suggested by these arguments and 
Eqs.(\ref{3}),(\ref{5}):
\bea
\label{6}
e^{- i V W(h, U) } = \sum_{n = - \infty}^{
 + \infty} \sum_{k=0}^{q-1} \exp \left\{ - \frac{i V}{4}
\left( h \, Log \, \frac{h}{2 e E} + 
\bar{h} \, Log \, \frac{ \bar{h}}{
2 e E } \right) \right. \nonumber \\ 
+ \left. i \pi V \left( k + \frac{q}{p} \,  
\frac{ \theta - i \log Det \, U + 2 
\pi n}{ 2 \pi} \right) \frac{h - \bar{h}}{
2 i} + \frac{i}{2}  V Tr( M U + h.c.) \right\} \;  ,
\eea
where $ M = diag (m_{i}  | \la \bar{\Psi}^{i} \Psi^{i} 
\ra | )$ 
and
the complex fields $ h , \bar{h} $ are defined as in Eq.(\ref{2})
with the substitution $ b_{YM} \rightarrow b = (11/3) N_c - (2/3)
 N_f $.
We note that all dimensional parameters in Eq.(\ref{6}) are fixed:
$ \la \bar{\Psi} \Psi \ra \simeq - (240 \, MeV)^3 , E = 
\la b \alpha_s /(32 \pi) G^2 \ra
\simeq 0.003 \, GeV^4 $ for $ N_c = N_f = 3 $ (see below), 
while the only unknown input are the integers $ p, q $ which 
are in general different from those standing
in Eq.(\ref{3}). 
They are related to a discrete symmetry surviving the anomaly,
and can be found only by a direct dynamical calculation. 
In particular, in SUSY theories there exists a special technique
\cite{SV} relating the strong coupling and weak coupling regimes,
which reveals that $ p = N_c , q = 1 $ for SU(N), $ p = N_c - 2,
q = 1 $ for SO(N), etc. We are not in a position to calculate these
numbers in QCD (see however \cite{3} which suggests  
$ q/p = 2 \xi = 8/(3b) $), and thus treat them as free parameters.
 We note that 
the ``dynamical" part of the anomalous effective potential (\ref{6})
can be written as $ W_d + W_{d}^{+} $ where 
\beq
\label{7}
W_{d} (h, U)  = \frac{1}{4} \frac{q}{p} h \, Log \left[ \left( 
\frac{h}{2 e E} \right)^{p/q}
\frac{ Det \, U }{ e^{-i\theta} } \right] -  \frac{1}{2}
Tr \, M U \; , 
\eeq
which is quite similar to  
the effective potential \cite{TVY} for SQCD \cite{Gomm}. 
%We stress 
%again 
%that our construction (\ref{6}) is based exclusively on WI's 
%of the theory. The holomorphic structure in Eq.(\ref{6}) is 
%a direct consequence of anomalous WI's. Below we will check that
%the anomalous WI's in QCD are reproduced from Eq.(\ref{6}).

We now wish to argue that Eq.(\ref{6}) represents the sought-after
anomalous effective Lagrangian realizing broken conformal and 
chiral symmetries of QCD. The following arguments will be suggested: 
(1) Eq.(\ref{6}) correctly reproduces the VVW ECL \cite{Wit2}
in the large $ N_c $ limit; (2) Eq.(\ref{6}) reproduces the anomalous 
conformal and chiral Ward identities of QCD; (3) it produces a 
new mass formula for the $ \eta' $ which appears reasonable 
phenomenologically; 
(4) it reproduces the known results for the $ \theta $ dependence 
at small $ \theta $, but may lead to a different behavior for 
larger values $ \theta > \pi/q $ if $ q \neq 1 $ as suggested by 
the method of \cite{3}. As a function of the phases $ \phi_i $,
the effective potential coincides at  $ N_c \rightarrow 
\infty $ 
with that 
of \cite{Wit2} for small $ \phi_i $,
but has a different form 
for larger values of  $ \phi_i $
in both cases $ q = 1 $ and $ q \neq 1 $. 
 
%last but not least, it suggests a more 
%attractive picture, with a smooth behavior in the quark masses, 
%of the physical $ \theta $ dependence in QCD
%than that of \cite{Wit2}, and matches the scenario of \cite{1} 
%for pure YM theory.

(1) The heavy ``glueball" fields $ h , \bar{h} $ can be integrated 
out in Eq.(\ref{6}) in the same way as was done in \cite{1}. The 
result is 
\bea
\label{8}
W_{eff}(U,U^{+}) =  - \lim_{V \rightarrow 
\infty} \; \frac{1}{V} \log \left\{ 
 \sum_{l} \exp \left[ 
V E \cos \left[ - \frac{q}{p} ( \theta - i \log  Det \, U ) 
+ \frac{2 \pi}{p}
\, l \right]  \right. \right. \nonumber \\
\left. \left. + 
\frac{1}{2} V \, Tr ( M U + M^{+} U^{+} ) \right] \right\}   
\; \; , \; \; l= 0,1, \ldots , p-1 
\eea
For small values of $ \theta 
 - i \log  Det \, U < \pi/ q $
the term with $ l = 0 $ dominates  the infinite volume limit. 
We obtain for this case
\beq
\label{9}
W_{eff}^{(l=0)}( U, U^{+}) = - E \cos \left[ - \frac{q}{p} ( \theta -
i \log Det \, U ) \right] - \frac{1}{2} \,  
Tr \, (MU + M^{+}U^{+} ) \; .
\eeq
Expanding the cosine (with the expansion parameter
$ q/p \sim 1/N_c $), we recover
exactly  the ECL of \cite{Wit2} at lowest order 
in $ 1/N_c $ (but only for $ \theta  -
i \log Det \, U < \pi/ q $), together with 
the ``cosmological" term $ - E = - 
 \la b \alpha_s /(32 \pi) G^2 \ra $ required by the conformal anomaly:
\beq
\label{VVW}
W_{eff}^{(l=0)}( U, U^{+}) = - E - \frac{ \la \nu^2 \ra_{YM} }{2} 
( \theta - i \log Det \, U )^2 - 
\frac{1}{2} \, Tr \, (MU + M^{+}U^{+} ) + \ldots \; , 
\eeq
where we used the fact \cite{1} that at large $ N_c $ $  E(q/p)^2 = - 
\la \nu^2 \ra_{YM} $ where $ \la \nu^2 \ra_{YM}< 0 $ is the 
topological susceptibility in pure YM theory. 
Corrections in $ 1/N_c $ stemming from Eq.(\ref{9}) 
constitute a new result.
Thus, in the large $ N_c $ limit 
the effective chiral potential (\ref{8}) coincides 
with that of \cite{Wit2} in the vicinity of the global minimum.
At the same time, terms with $ l \neq 0 $ in Eq.(\ref{8}) result 
in different global properties of the effective chiral potential
in comparison with the one of Ref.\cite{Wit2}, see below.
% We thus see that the small $ \theta $ 
%behavior of the theory is correctly described by the 
%VVW potential
%(\ref{VVW}). At the same time, for large $ \theta > \pi/q $ 
%branches with $ l \neq 0 $ in Eq.(\ref{8}), which can not be seen
%n the VVW formula (\ref{VVW}), become important. As will be discussed 
%below, this changes completely the global picture of physical $ 
%\theta $ dependence
%in comparison with the VVW scenario \cite{Wit2}.  

(2) It is easy to check that the anomalous chiral and conformal 
Ward identities (WI's) are reproduced by Eq.(\ref{6}). The 
anomalous chiral WI's are automatically satisfied by the 
substitution (\ref{5}) for any $ N_c $, in accord with \cite{Wit2}.
Further, it can be seen that the anomalous conformal WI's of 
\cite{NSVZ} for zero momentum correlation functions of operator $ G^2 $
are also satisfied with the above choice of constant $ E $. As 
an important example, let us 
calculate the topological susceptibility 
in QCD near the chiral limit from
Eq.(\ref{8}). For simplicity, we consider the limit of $ SU(N_f) $ 
isospin symmetry with $ N_f $ light quarks, $ m_{i} \ll \Lambda_{QCD} $.
For the vacuum energy for small $ \theta < \pi/q $ we obtain (see
Eq.(\ref{18}) below)
\beq
\label{10}
 E_{vac} (\theta) = -E  + m \la \bar{ \Psi} \Psi \ra  N_{f}
\cos \left( \frac{\theta}{N_{f}} \right) + O(m_{q}^2)  \; . 
\eeq
Differentiating this expression twice in $ \theta $, we reproduce
the result of \cite{SVZ}:
\beq
\label{11}
\lim_{ q \rightarrow 0} \; 
i \int dx \, e^{iqx} \lo T \left\{ \frac{\alpha_s}{8 \pi} 
G \tilde{G} (x)  \, 
\frac{\alpha_s}{8 \pi} G \tilde{G} (0) \right\}  \ro =  
 \frac{1}{N_f} m \la \bar{ \Psi} \Psi \ra  + O(m_{q}^2) \; .
\eeq

(3) To find a mass formula for the $ \eta' $ meson, we should 
calculate the matrix of second derivatives at the 
minimum of the effective potential for small $ \theta $. 
Neglecting for simplicity a small $ \pi^{0} - \eta - \eta' $ 
mixing, we obtain
\beq
\label{12}
f_{\eta'}^{2} \, m_{\eta'}^2  = \frac{8}{9 b} N_{f} 
\la \frac{ \alpha_s}{\pi} \, G^2 \ra - \frac{4}{N_{f}} 
\sum_{u,d,s} m_{i} \la \bar{\Psi}_{i} \Psi_{i} \ra + O(m_{q}^{2}) \; . 
\eeq
(Here we used the value $ q/p = 8/(3b) $ suggested by the method of 
\cite{3}. The choice $ p = N_c , q = 1 $ would produce a numerically 
close result.)
This mass relation for the $ \eta' $ appears reasonable 
phenomenologically. 
Note that, according to Eq.(\ref{12}), the strange 
quark contributes 30-40 \% of the $ \eta' $ mass. In the formal
limit $ N_c \rightarrow \infty, m_q \rightarrow 0 $ 
Eq.(\ref{12}) coincides with
the relation obtained in \cite{2}. Eq.(\ref{12}) suggests
that the main origin of the large mass of the $ \eta' $ is 
the conformal anomaly in QCD, in accord with arguments of \cite{2,1}.
 
(4) The mail difference of the 
effective chiral potential (\ref{8}) from that of \cite{Wit2}
is its non-analytic structure in the limit
$ V \rightarrow \infty $, with cusp singularities 
at certain values of  $ \theta -  i \log  Det \, U  $.
These cusp singularities are analogous to the ones arising in 
the case of pure gluodynamics \cite{1} for the vacuum energy as 
function of $ \theta $, showing non-analyticity
of the $ \theta $ dependence at certain values of $ \theta $.
In the present case, the effective potential for the light 
chiral fields analogously becomes non-analytic at some 
values of the fields. The origin of this non-analyticity 
is the same as in the pure YM case - it appears when the topological
charge quantization is imposed explicitly at the effective 
Lagrangian level. Thus, the cusp structure of the 
effective potential seems to be an unavoidable consequence of the 
topological charge quantization (which was not explicitly imposed 
in \cite{Wit2}). This fact was first noted in the 
context of SUSY theories \cite{KS,KKS}, where similar formulas 
arise.

For practical applications, it is convenient to describe the 
non-analytic effective
chiral potential (\ref{8}) by a set of analytic functions
defined on different intervals of the combination
$ \theta - i \log Det \; U $. Thus, in   
the infinite volume limit the effective potential for the 
fields $ U= diag( \exp i \phi_q ) $ 
is dominated by its $l $-th branch:
\beq
\label{13}
W_{eff}^{(l)} = - E \cos \left( - \frac{q}{p} \theta + 
\frac{q}{p} \sum \phi_{i} + \frac{2 \pi}{p} \, l \right) 
- \sum M_{i} \cos \phi_i  \; \; , \; \;   l= 0,1, \ldots , p-1
\eeq
if 
\beq
\label{14}
(2 l - 1) \frac{\pi}{q} \leq \theta - \sum \phi_i < (2 l + 1) 
 \frac{\pi}{q} \; \; .
\eeq
This can be viewed as the set of $ ``p" $ different effective potentials 
describing different branches in Eq.(\ref{6}). The 
periodicity in $ \theta $ is realized on the set of 
potentials (\ref{13}) as a whole, precisely as it occurs in the 
pure gauge case \cite{1} where different branches undergo a
cyclic permutation under the shift $ \theta \rightarrow \theta - 2 \pi $. 
As is seen from Eq.(\ref{13}), the shift 
 $ \theta \rightarrow \theta - 2 \pi $ transforms the branch with
$ l = k $ into the branch with $ l = k + q $. 
In addition, as long as $ q \neq 1 $, there exists another 
series of cyclic permutations
corresponding to $ l = k $ and $ l = k+1 $ in the above 
set, which are related to each 
other by the shift $ \theta \rightarrow \theta - 2 \pi/q $.
If the number $ q $ were $ q = 1 $, the 
two series would be, of course, the same.
As there exist certain arguments in favor of the choice $ q \neq 1 $
\cite{3}, we consider both cases.
%Thus, a value $ q \neq 1 $ implies some discrete symmetry arising
%at the quantum level. As we mentioned earlier, the approach of 
%\cite{3} suggests that $ q = 8 $ for both YM theory and QCD. 
%Although we do not see any explicit flaws in the method exploited 
%in \cite{3} to find the parameter $ q $, the problem of its 
%precise value may deserve a further study. 
As we will argue below, 
some properties of the theory are very different depending on whether
$ q = 1 $ or $ q \neq 1 $.

The minimization equations stemming from Eqs.(\ref{13}),(\ref{14})
can be analysed numerically or analytically in different limits, as
was done in \cite{Wit2}. For the case $ q = 1 $, we find for the 
locations and number of vacua and the $ \theta $ dependence 
of the vacuum energy the same results as those of \cite{Wit2}. 
In particular, we  find the 
vacuum doubling at $ \theta = \pi $ (Dashen's phenomenon \cite{Dash}) 
if $ 
m_{u} m_{d} > m_{s} | m_{d} - m_{u} |  $.
As this condition is not satisfied with the physical values of the 
quark masses, we conclude that  in the case $ q = 1 $ there is 
the unique vacuum for all values of $ \theta $. The only difference 
of our results from those of \cite{Wit2} for $ q = 1 $ is a different 
global form of the effective chiral potential, with cusp singularities 
at certain values of the fields. No metastable local minima  
of the effective potential exist for any 
$ \theta $ (they appear only for $ N_f > 4 $), though 
saddle points are there.

If, on the other hand, $ q \neq 1 $, we find the results 
for the $ \theta $ dependence and global structure of the 
effective chiral potential which are very different from 
those of Ref.\cite{Wit2}. In particular,
in 
the case of equal masses $ m_{i} = m \ll \Lambda_{QCD} $ the 
lowest energy state is described by
\beq
\label{18}
\phi_{i}^{(l=0)} = \frac{\theta}{N_{f}} \; , \;   E_{vac} 
(\theta) = - E - M  N_{f} 
\cos \left( \frac{\theta}{N_{f}} \right) + O(m_{q}^2)  \; , 
\; \theta
\leq \frac{\pi}{q} \; ;  
\eeq
\beq
\label{19}
\phi_{i}^{(l=1)} = \frac{\theta}{N_{f}} - \frac{2 \pi}{q N_f}
\; , \;   E_{vac} (\theta) =  - E - M N_{f} 
\cos \left( \frac{\theta}{N_{f}}- \frac{2 \pi}{q N_f} 
\right) + O(m_{q}^2)  \; , \; \frac{\pi}{q} \leq \theta
\leq \frac{3 \pi}{q} \; ; 
\eeq
{\it etc}. Thus, the solution (\ref{18}) coincides with the 
one obtained 
by VVW \cite{Wit2} at small $ \theta < \pi/ q $ up to $ O(m_{q}^2) $ 
terms. 
However, at larger values of $ \theta $ the true vacuum switches 
from (\ref{18}) to (\ref{19}) with the cusp singularity at 
$ \theta = \pi/q $. (If $ q = 1 $, the results of \cite{Wit2}
are recovered). The interesting feature of the case $ q \neq 1 $ is 
that the vacuum doubling at the points
\beq
\label{17}
\theta_k = (2 k + 1) \, \frac{\pi}{q} \; \; , \; \;
 k = 0, 1, \ldots, 
p-1 
\eeq
holds irrespective of the values of the light quark masses.
This can be seen from the 
fact that
the equations of motion for any two branches with 
$ l = k $ and $ l = k+1 $ from the set (\ref{13}) are related 
by the shift $ \theta \rightarrow \theta - 2 \pi/ q $. Thus, the 
extreme sensitivity of the theory to the values of the light
quark masses in the vicinity
of the critical point in $ \theta $ is avoided in our scenario
if $ q \neq 1 $, while the location of the critical point is 
given by $ \theta_c = \pi/q $ instead of the ``standard" $ \theta_c =
\pi $. Another interesting feature of the scenario $ q \neq 1 $ 
is the appearance of metastable vacua which exist for any value 
of $ \theta $, including $ \theta = 0 $. For the physical 
values of the quark masses, we find $ q - 1 $ additional local
minima of the effective chiral potential, which are separated 
by barriers from the true physical vacuum of lowest energy. 
If true (i.e. if $ q \neq 1 $), this result can be important for 
a number of physical problems. A discussion of these 
issues, as well as a more detailed derivation of the results 
presented in this letter will be given elsewhere.

%{\bf 4.} In summary, we have presented arguments in favor of correctness
%of Eq.(\ref{6}) as the anomalous effective potential for QCD.
%$The most important result that follows is a new 
%scenario of the $ \theta $ dependence in QCD which is very much 
%the same as in pure YM theory, and does not change with the light 
%quark masses as long as $ m_u, m_d , m_s < \Lambda_{QCD}$. 
%A more detailed analysis and phenomenological applications will
%be given elsewhere.  

\end{document}